# Global to local impacts on atmospheric $CO_2$ caused by COVID-19 lockdown

Ning Zeng[1,2,*], Pengfei Han[3], Di Liu[3], Zhiqiang Liu[3], Tomohiro Oda[4,5,1], Cory Martin[6], Zhu Liu[7], Bo Yao[8], Wanqi Sun[8], Pucai Wang[9], Qixiang Cai[3], Russell Dickerson[1,2], Shamil Maksyutov[10]

[1] Department of Atmospheric and Oceanic Science, University of Maryland, USA
[2] Earth System Science Interdisciplinary Center, University of Maryland, USA
[3] Laboratory of Numerical Modeling for Atmospheric Sciences & Geophysical Fluid Dynamics, Institute of Atmospheric Physics, Chinese Academy of Sciences, Beijing, China
[4] Universities Space Research Association, Columbia, MD, USA
[5] NASA Goddard Space Flight Center, Greenbelt, MD, USA
[6] RedLine Performance Solutions, LLC and National Weather Service of National Oceanic and Atmospheric Administration, USA
[7] Department of Earth System Science, Tsinghua University, Beijing 100084, China
[8] Meteorological Observation Centre, China Meteorological Administration, Beijing, China
[9] Laboratory for Middle Atmosphere and Global Environment Observation, Institute of Atmospheric Physics, Chinese Academy of Sciences, Beijing, China
[10] National Institute for Environmental Studies, Tsukuba, Japan
[*] Corresponding author email: zeng@umd.edu

## Abstract

**The world-wide lockdown in response to the COVID-19 pandemic in year 2020 led to economic slowdown and large reduction of fossil fuel $CO_2$ emissions [1,2], but it is unclear how much it would reduce atmospheric $CO_2$ concentration, the main driver of climate change, and whether it can be observed. We estimated that a 7.9% reduction in emissions for 4 months would result in a 0.25 ppm decrease in the Northern Hemisphere $CO_2$, an increment that is within the capability of current $CO_2$ analyzers, but is a few times smaller than natural $CO_2$ variabilities caused by weather and the biosphere such as El Nino. We used a state-of-the-art atmospheric transport model to simulate $CO_2$, driven by a new daily fossil fuel emissions dataset and hourly biospheric fluxes from a carbon cycle model forced with observed climate variability. Our results show a 0.13 ppm decrease in atmospheric column $CO_2$ anomaly averaged over 50S-50N for the period February-April 2020 relative to a 10-year climatology. A similar decrease was observed by the carbon satellite GOSAT[3]. Using model sensitivity experiments, we further found that COVID, the biosphere and weather contributed 54%, 23%, and 23% respectively. In May 2020, the $CO_2$ anomaly continued to decrease and was 0.36 ppm below climatology, mostly due to the COVID reduction and a biosphere that turned from a relative carbon source to carbon sink, while weather impact fluctuated. This seemingly small change stands out as the largest sub-annual anomaly in the last 10 years. Measurements at marine boundary layer stations such as Cape Kumukahi, Hawaii exhibit 1-2 ppm anomalies, mostly due to weather and biosphere. At city scale, on-road $CO_2$ enhancement measured in Beijing shows reduction of 20-30 ppm, consistent with drastically reduced traffic during the lockdown, while station data suggest that the expected COVID signal of 5-10 ppm was swamped by weather-driven variability on multi-day time scales. By contrast, a clear stepwise drop of 10-20 ppm at the city-wide lockdown was observed in the city of Chengdu where weather is less variable. The ability of our current carbon monitoring systems in detecting the small and short-lasting COVID signal on the background of fossil fuel $CO_2$ accumulated over the last two centuries is encouraging. The COVID-19 pandemic is an unintended experiment whose impact suggests that to keep atmospheric $CO_2$ at a climate-safe level will require sustained effort of similar magnitude and improved accuracy and expanded spatiotemporal coverage of our monitoring systems.**

The unprecedented world-wide lockdown in the first few months of year 2020 led to wide-spread reduced economic activities. As a result, fossil fuel $CO_2$ emissions were reduced by 7.9% in the first 4 months of 2020 due to reduced transportation, industrial and power generation [2] and anticipated annual reduction of 4-7% [1]. This level of emissions reduction is on par with the annual target set out by the Paris climate agreement, but the COVID-19 induced reduction is short-lived as economic activities rebounds subsequently. While the lockdown increased activities such as tele-collaboration that benefit climate, other measures do not lead to transformation needed in energy systems. Monitoring and understanding such processes from global to local scales are of critical importance for achieving our climate goals. Over the last few decades, the scientific community has been developing world-wide carbon monitoring and information systems aiming at enabling the monitoring and verification of emissions reduction goals [4-7].

**How big is the impact on atmospheric $CO_2$ of this dramatic but short-term reduction in fossil fuel emissions?**

A back-of-the-envelope calculation goes as following. Current fossil fuel emissions rate is 10 GtC $y^{-1}$ (Gigatonne carbon per year), of which about half is taken up by carbon sinks on land and in the ocean, with the remaining half left in the atmosphere, resulting in a $CO_2$ increase of 2.5 ppm per year, as observed in a world-wide network of $CO_2$ observatories such as the renowned Mauna Loa station in Hawaii. Assuming 7% reduction or 0.7 GtC for the whole year of 2020 (high estimate of reference [1]), it would cause a 0.18 ppm relative decrease in global atmospheric $CO_2$ at the end of the year.

In reality, the emissions reduction does not occur uniformly throughout the year, for example, China in February and Europe, US and India in March-April (Fig. S1). The estimated reduction of 7.9% in January-April 2020 [2] corresponds to a decrease of 0.26 GtC, a rather small quantity for atmospheric $CO_2$. However, we expect the COVID signal to stay largely in the Northern Hemisphere (NH) for these few months because atmosphere inter-hemispheric transport takes approximately one year. We further assume that the carbon sinks have not kicked in because of dormant winter vegetation and sluggish ocean-atmosphere gas exchange. We therefore expect a 0.25 ppm decrease of Northern Hemisphere $CO_2$ at the end of April. This magnitude of change is within the capability of today's high-accuracy $CO_2$ analyzers, but small for carbon satellites such as NASA's OCO-2 and the Japanese GOSAT with targeted precision of ~1 ppm and their ground calibration of 0.4 ppm[3,8,9].

Challenges also arise from the fact that besides fossil fuel emissions, atmospheric $CO_2$ is also strongly influenced by the changes in the biosphere and atmospheric transport (weather).
The growth and decay of the northern vegetation causes a seasonal $CO_2$ amplitude of 6 ppm, while interannual climate variability such as ENSO causes biogenic $CO_2$ changes of 1-3 ppm[10-13]. Thus, a key question is whether a 0.25 ppm COVID signal can be seen among other (natural) variabilities. We explored this question with carbon cycle models and a suite of space-borne and ground-based observations.

## Modeling atmospheric $CO_2$ response to emissions reduction and biospheric anomalies

First, we created a high spatial-temporal resolution fossil fuel emissions ($F_{FE}$) dataset with daily emissions data from a near real time product (updated from Liu et al., (2020)[2]; see Method). The daily country-level data was disaggregated to model grid resolution based on the ODIAC emissions database [14]. As seen in Fig. 1a (detailed temporal evolution in Fig. S1-S2), carbon emissions intensity decreased by more than 5 gC $m^{-2}$ $mo^{-1}$ during February-April 2020 in East Asia, Europe, US and India, relative to the same period in 2019. While consistent with the temporal variations in the original country-level data [2], the spatially disaggregated data here further shows emissions reductions concentrated in industrialized regions and areas with high population density such as the North China Plain, India's Ganges Basin, Northeast and Midwest US, as well as isolated centers such as Sao Paolo.

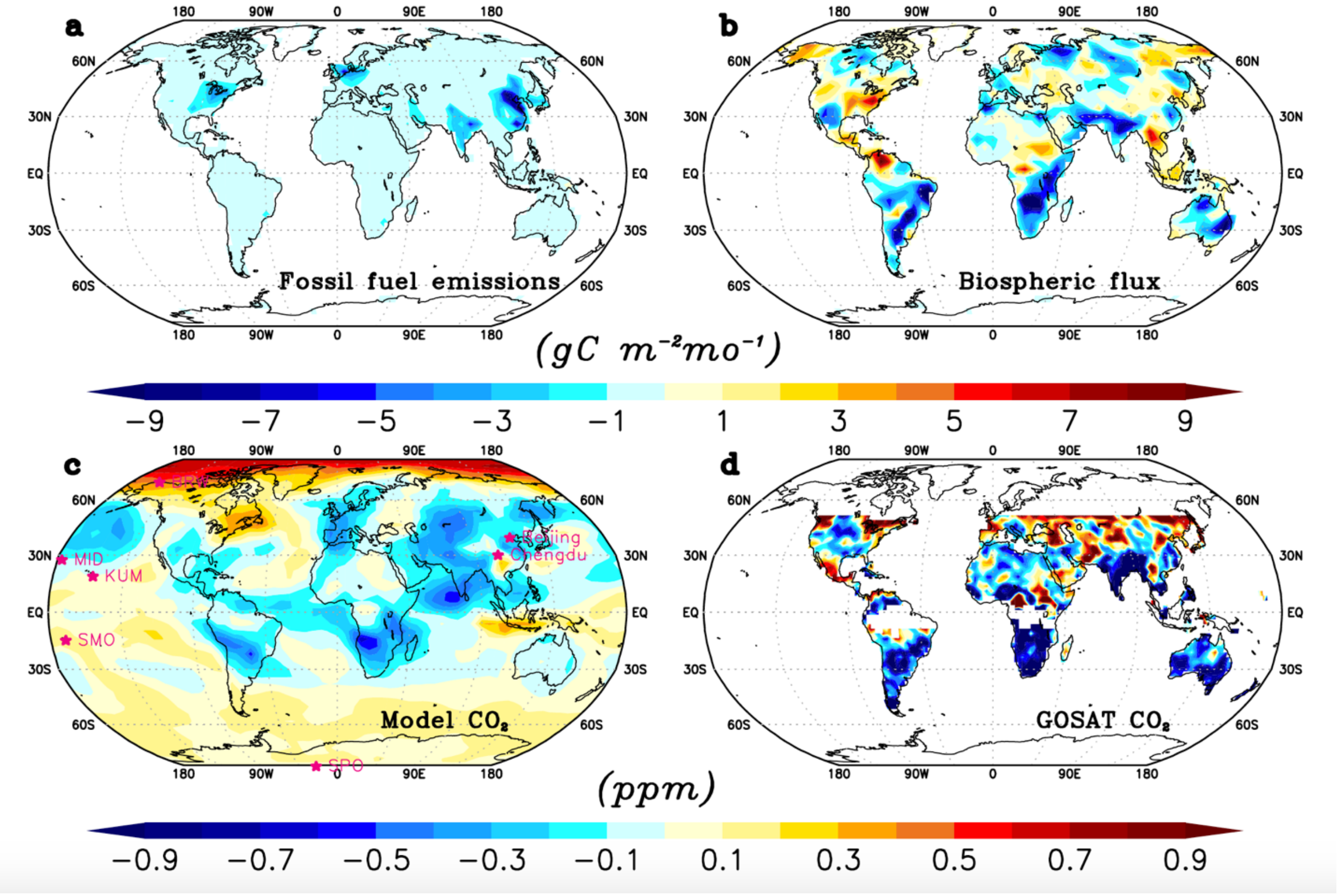

*Figure 1. Anomalies for the period of February-April 2020 of (a) Fossil fuel emissions ($F_{FE}$); (b) terrestrial biosphere-atmosphere flux ($F_{TA}$); (c) atmosphere transport model simulated column $CO_2$ (vertically averaged); (d) Observed column $CO_2$ from GOSAT. Anomalies in (a) are relative to 2019 while those in (b)-(d) are relative to climatology of 2010-2019. Fluxes are in gC $m^{-2}$ $mo^{-1}$, and $CO_2$ in ppm. Locations in (c) are selected ground $CO_2$ observation stations for model-data comparison.*

For terrestrial biosphere-atmosphere flux ($F_{TA}$), we used a dynamic vegetation and terrestrial carbon cycle model VEGAS [10,15], while ocean-atmospheric flux came from a multi-model product (see Method). The biospheric fluxes (Fig. 1b) have regional variations often larger than $F_{FE}$ reduction, driven by climate variability. Overall, the terrestrial biosphere had wide-spread negative anomalies that was particularly strong in April (Fig. S3). As a result, the spatial pattern of net flux $F_{net}$ was dominated by biospheric fluxes.

Interestingly, when summed up globally, the magnitude of 2020 $F_{FE}$ anomaly relative to 10-year climatology is comparable to that of $F_{TA}$ (Fig. 2a), each about 50 MtC mo$^{-1}$ decrease by April 2020. This is because fossil fuel emissions are reduced almost everywhere during COVID-19 lockdown, whereas the positive and negative anomalies in $F_{TA}$ tend to cancel out.

Consequently, the total flux anomaly $F_{net}$ reaches -100 MtC mo$^{-1}$.

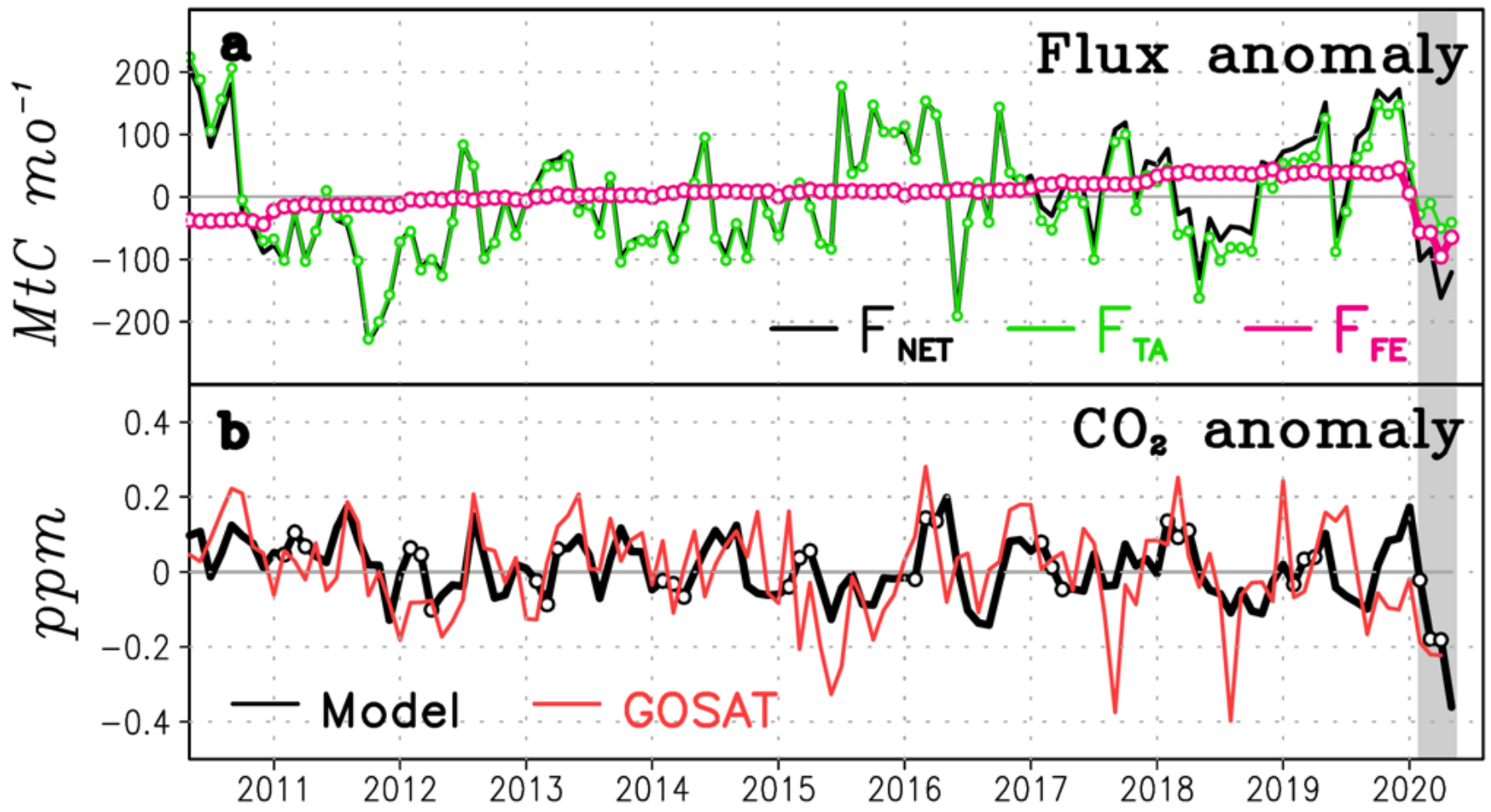

*Figure 2. (a) Anomalies of global total Fossil Fuel Emissions ($F_{FE}$), terrestrial biosphere-atmosphere flux ($F_{TA}$), and net fluxes ($F_{net}$) relative to a 10 year climatology; (b) Detrended anomalies of model simulated column $CO_2$, with closed black circles marking each year's February, March and April, while red line is the same but for GOSAT satellite column $CO_2$ data, both averaged for land area between 50S-50N to make them as comparable as possible. Vertical shading of February-May 2020 indicates the 'COVID-19 period'.*

These fluxes were then used to drive the community GEOS-Chem atmospheric transport model for the period of January 2008-May 2020 (see Method). The output of the model is a 4-dimensional depiction of spatial-temporal evolution of atmospheric $CO_2$ that can be compared to various types of $CO_2$ observations, as well as expected COVID impact (Fig. S4-S7). Using a method termed DCA (Detrended Climatology and Anomaly; see Method), in which a $CO_2$ time series is detrended to remove the low-frequency signal and climatological seasonal cycle, we calculated sub-annual anomalies of column $CO_2$, shown in Fig. 2b. After a large positive anomaly in Jan 2020, the $CO_2$ anomaly trended downward, and was 0.19 ppm lower than long-term climatology in March-April 2020, and 0.36 ppm lower in May (Fig. 2b shaded period). This anomaly is twice the second lowest monthly value in July 2016 and stands out among the past 10 years, and particularly so for the same season.

**Observations from carbon satellite**

The Japanese GOSAT carbon satellite, launched in 2009, has collected column $CO_2$ ($XCO_2$) data for over 10 years. The spatial pattern of February-April 2020 GOSAT anomalies (Fig. 1d) are similar to the model over India, southern Africa, South America and central US where large negative values are wide-spread, though the overall spatial correlation is modest and significant sign differences exist near the northern edge in Eurasia (Fig. 1 c and d). Detailed monthly evolution shows similar broad patterns of agreement as well as larger detailed differences (Fig. S8 vs. S9). In a sense, the differences are expected because of satellite's sparse spatial-temporal sampling, particularly at higher latitudes and cloudy regions. However, the signal-to-noise ratio is enhanced via spatial-temporal averaging: the time series of $CO_2$ anomalies averaged over 50S-50N (Fig. 2b) shows reasonable agreement. Most encouragingly, the drop in 2020 stands out both in model and GOSAT.

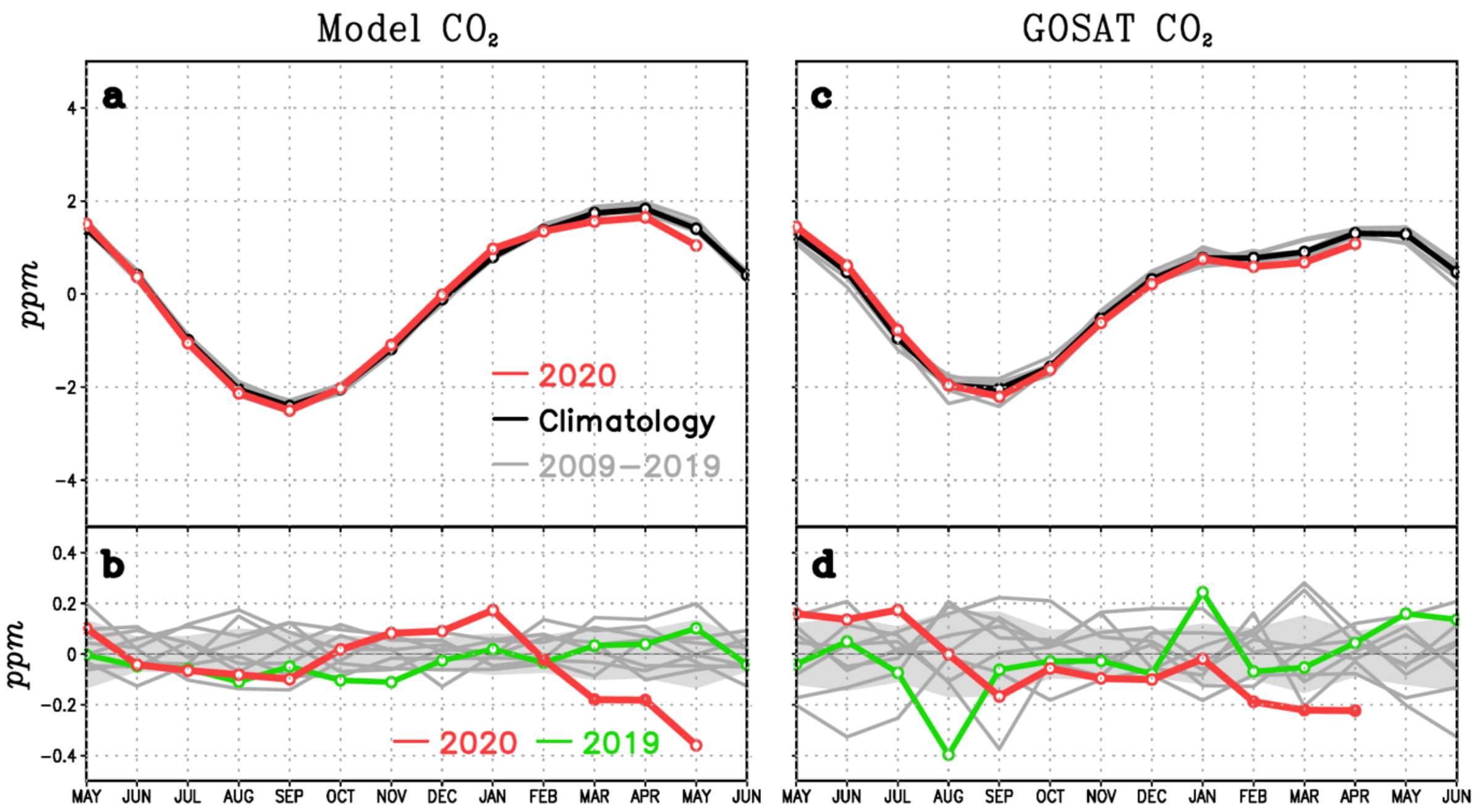

*Figure 3. (a) Seasonal cycles of detrended modeled column $CO_2$ averaged over 50S-50N land region, with each individual year plotted in gray lines, climatological mean of the 10 years of 2009-2019 in black, and most recent year May 2019 - April 2020 in red; (b) $CO_2$ anomalies relative to the climatology in (a) with same color scheme, and green for the previous year (May 2018 – April 2019); (c)-(d) As in (a)-(b) but for the GOSAT satellite observed column $CO_2$.*

To better appreciate how unusual year 2020 was, we plotted the $CO_2$ seasonal cycle from May 2019 to April 2020 against those of the previous 10 years (Fig. 3). The 50S-50N land average $CO_2$ for especially March-April is outside the standard deviation as well as the envelope of the previous 10 individual years for both model and GOSAT. Such agreement lends confidence in using both model and satellite column $CO_2$ for short-term anomaly detection. We also analyzed data at smaller scales including the NH 0-50N where strongest COVID-signal is expected. The model shows clear negative anomaly but the corresponding GOSAT signal is not as statistically significant as in 50S-50N (Fig. S10), likely due to increase in noise towards smaller scales. The more robust 50S-50N signal has contribution from GOSAT's ability to capture the anomalies in the Southern Hemisphere which is largely a biospheric signal. Similarly, a large decrease in GOSAT anomaly for Eastern China in February 2020 (Fig. S11), though may be partly related to COVID reduction, is not statistically robust.

**Attribution of the 2020 $CO_2$ drawdown to the biosphere, weather, and COVID reduction**

We conducted two additional model sensitivity experiments to delineate the effects of biosphere, atmospheric transport (weather), and COVID reduction in $F_{FE}$ (See Method and aspects of the detailed result in Fig. S4-S7). The monthly evolution of $CO_2$ anomalies from these experiments (Fig. 4) indicate that the roles of the biosphere, weather, and COVID vary from month to month. In February 2020, biosphere has a near zero but positive anomaly, while COVID effect steadily increases from February to May. The weather effect was large in March and May to drawdown $CO_2$, but small in both February and April, all in comparison with 2019 by model design. Averaged over February-April, the biosphere contributed 0.030, weather 0.029, and COVID 0.069 ppm, leading to a February-April average of 0.128 ppm $CO_2$ drawdown with March and April both about 0.18 ppm. By percentage, the biosphere, weather, and COVID contributed 23%, 23%, and 54%, respectively. In May 2020, all three factors contributed to $CO_2$ drawdown, leading to a rapid decrease to 0.36 ppm.

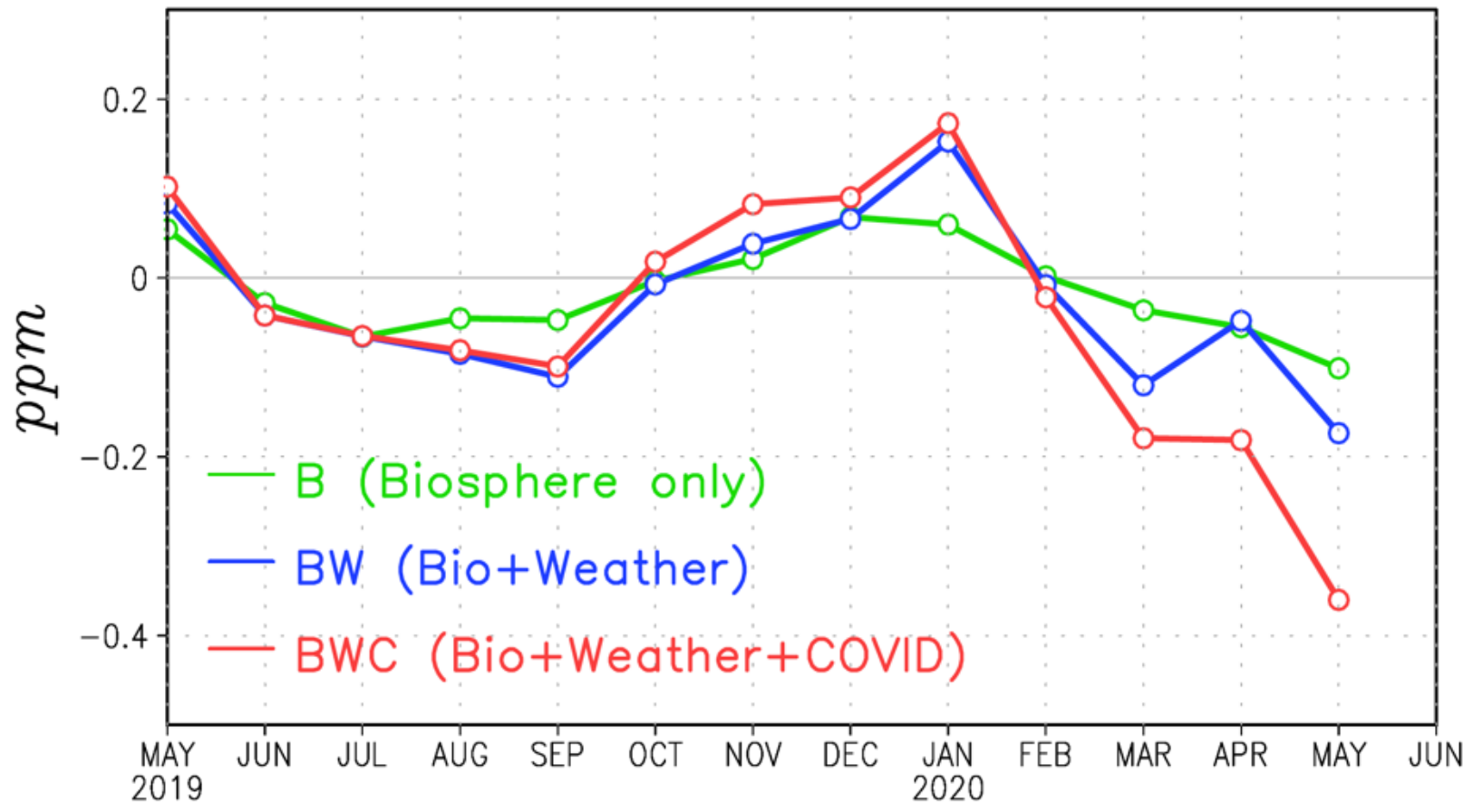

*Figure 4. Model sensitivity experiments to separate the effects from the 3 factors: Biospheric flux (B), atmosphere transport (Weather or W), and COVID-19 induced reduction in fossil fuel emissions (C). Starting from the original experiment that has all three effects of Biosphere, Weather and COVID (BWC), Experiment BW removes COVID impact, while Experiment B removes both COVID and weather impact. By experimental design, biospheric effect is relative to a 10-year climatology while COVID and weather effects are relative to 2019 (see Method). Shown are cumulative contribution to mean column $CO_2$ anomalies over 50S-50N. The difference of two experiments represents contribution from an individual factor.*

In summary, the three factors impacted $CO_2$ in an intricate way. In the period February-May 2020, the biosphere was recovering from a large carbon loss (relative to climatology) from year 2019, mostly in South America, Africa, and Australia in response to a moderate El Nino[10], with additional $CO_2$ uptake from Northern Hemisphere including South Asia and Siberia (Fig. S3). COVID emissions reduction was the most spatiotemporally consistent factor, contributing to majority of the $CO_2$ decrease. The weather impact on $CO_2$ fluctuated from month to month. The weather effect is not predominant because its variability tends to be averaged out on global scales. Its importance rises at smaller scales to which we turn our attention now.

**$CO_2$ changes at surface atmosphere background stations**

Similar to the global-scale analysis above, we also analyzed data from atmospheric background $CO_2$ stations from the NOAA global reference network ObsPack [16] (see Method). Fig. 5 compares model with surface observation at five marine boundary layer stations that span the remote Pacific region from North to South. Flask sampling at these sites is carefully conducted to represent atmospheric background concentrations on the scale of hundreds to thousands of kilometers. Among these stations is Cape Kumukahi, Hawaii (KUM), a site downslope of the Mauna Loa observatory, but is better than Mauna Loa at capturing large-scale atmosphere background signal and more comparable with model which does not resolve the island.

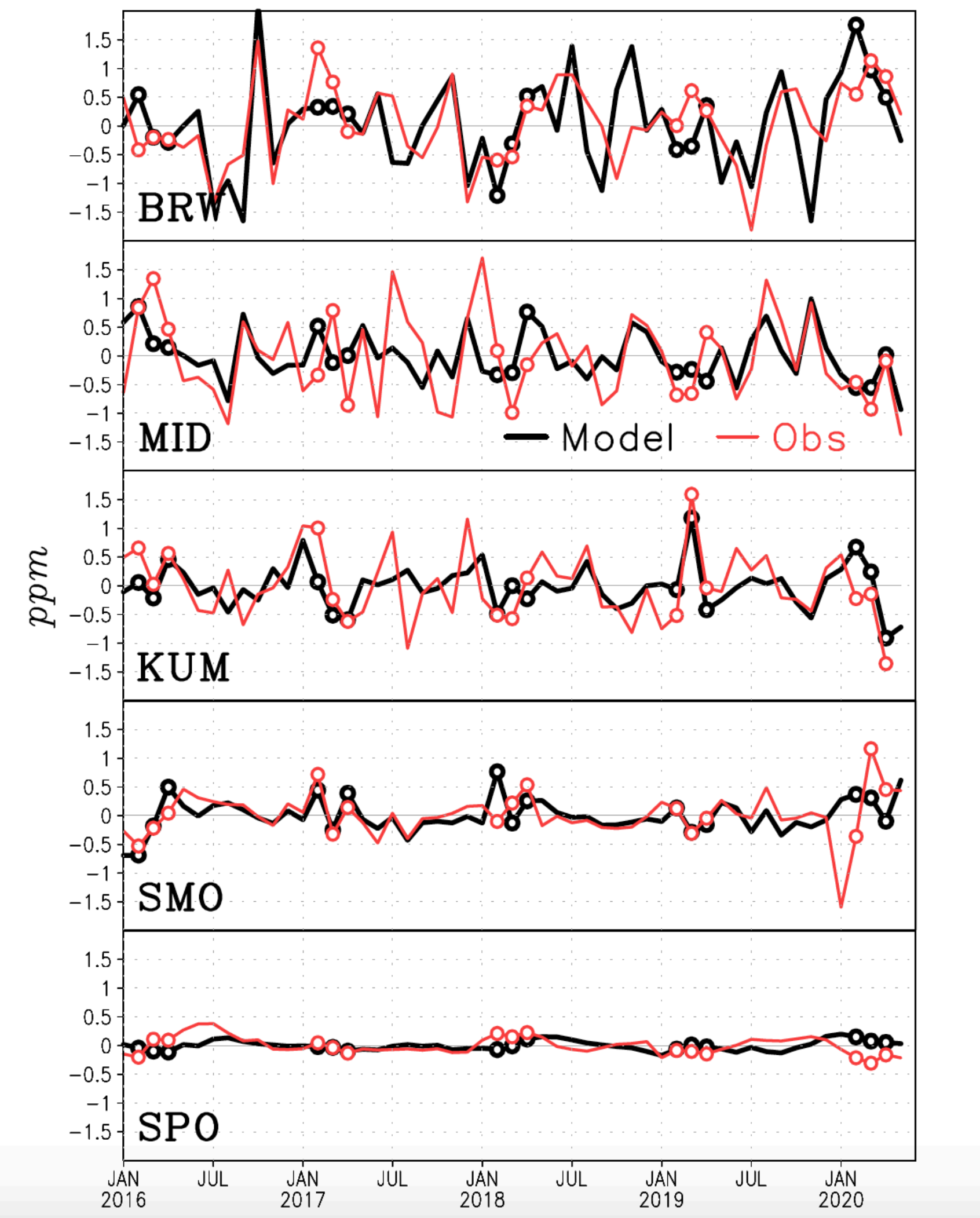

*Figure 5. Model-data comparison of sub-annual $CO_2$ anomalies (ppm) at 5 atmospheric background sites: Barrow, Alaska (BRW); Midway Island, North Pacific Ocean (MID); Cape Kumukahi, Hawaii (KUM); American Samoa in South Pacific Ocean (SMO), South Pole Station (SPO). Site locations are marked in Fig. 1c. Similar to Fig. 2b, model in black, observation in red, with filled markers indicating the months of Feb-Apr for each year. Data are shown only for the recent few years while anomalies are relative to a 10 year climatology.*

The anomalies at Kumukahi are relatively small in February and March, 2020, but decrease rapidly to lower than -1 ppm in April. Model and observation are broadly similar, in particular, the decrease in April. It is tempting to associate this drop with COVID reduction. However, a closer look at model sensitivity experiments (Fig. S12) reveals large month-to-month fluctuations from biosphere and weather, for example, a positive anomaly up to +1 ppm from the biosphere and a similarly large negative anomaly from weather in March. Of the total $CO_2$ drop of 1 ppm in April, 0.5 comes from biosphere, 0.5 from weather, and only 0.2 ppm from COVID.

Unlike a decrease in Hawaii, $CO_2$ at Barrow, Alaska (BRW) shows an increase in early 2020, while Midway Island in the North Pacific Ocean (MID) has a decrease in February-March and rebound in April, both attributable to the biosphere and/or weather (Fig. S12). Thus, while the sub-annual signal of 1-2 ppm at these marine surface stations is a few times larger than global mean column anomalies seen by the satellite, the variability at a given station is dominated by the biosphere and weather. Nevertheless, the 0.1~0.2 ppm decrease due to COVID-19, though smaller than biospheric and weather effect, are separable using model, and is consistent with the global-scale COVID induced drop discussed above. The overall consistency between model and station observations suggests the ability of both model and observation in capturing these sub-annual changes, regardless of the origin, thus lending support for interpreting the signal using model experiments.

## City-scale $CO_2$ changes

Because major fraction of emissions comes from cities with high human activities, one can expect large COVID signal in urban $CO_2$ data. Thus, we analyzed $CO_2$ measurements in Beijing and Chengdu.

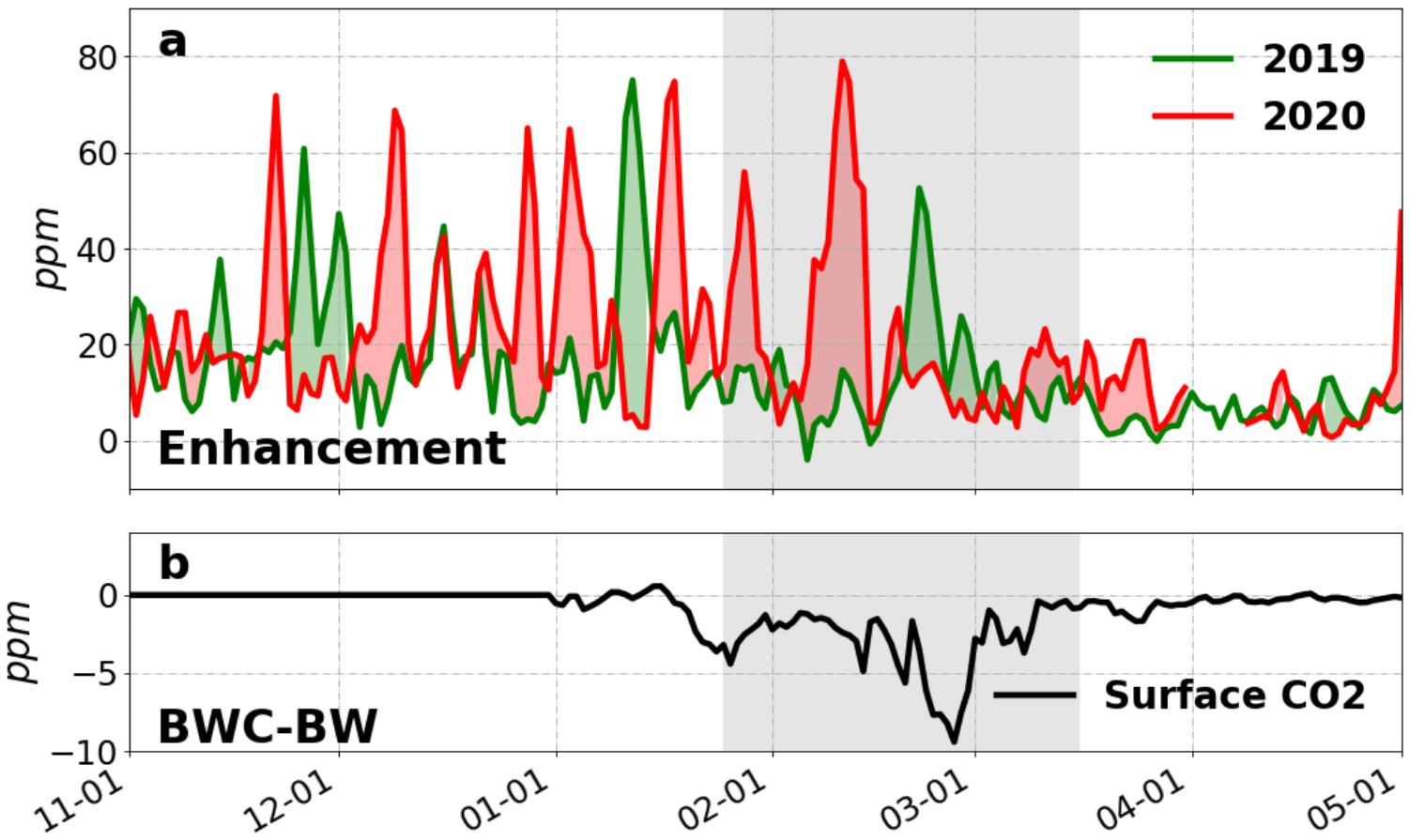

*Figure 6. Daily $CO_2$ measured in Beijing. (a) Measured $CO_2$ enhancement (Xianghe station (a near-city site) minus Xinglong, a rural site in the mountains northeast of Beijing, for the period Nov2019-Apr2020 (labeled as 2020), compared to the year before (Nov2018-Apr2019, labeled as 2019); (b) Model simulated $CO_2$ difference caused by COVID-19 emissions reduction (Experiment BWC – Experiment BW). Vertical shading indicates the lockdown period in Beijing.*

Surface observations in Beijing shows $CO_2$ for the winter-spring (December 2019-April 2020) compared to the same period the year before (Fig. 6a). During the pre-COVID period of December-January, $CO_2$ is significantly higher in 2020 than the year before, because this winter's atmosphere was less 'ventilated'. February was dominated by two high-$CO_2$ weather events, one in each year. The expected low $CO_2$ values due to COVID in January-February (Fig. 6b) 'predicted' by the model do not have a clear correspondence in the weather-dominated $CO_2$ fluctuations, and in general, the modeled magnitude of change is much smaller than the variability. During March-April, the difference between the two years decreases to much smaller values. It is tempting to explain this difference as the result of emissions reduction, but it is mostly brought about by weather regime shift in the spring season with dominantly northwesterly wind from Mongolian Plateau. Thus, although COVID-19 signal is large, it is 'buried' in even larger weather noise. Other cities are similarly dominated by weather (Fig. S13), for example, a monthly drop of 1-2 ppm at New York City and Delhi, an increase of 1-2 ppm in Washington DC and Paris.

Interestingly, $CO_2$ measured at the city of Chengdu shows a stepwise drop on January 24, the day before traditional Chinese Lunar New Year, followed by city-wide lockdown with little urban activity for the next 1-2 months (Fig. 7). The difference between the month before and after the lockdown is 20-30 ppm and the temporal characteristics is consistent with a COVID signal. Concurrent particulate matter (PM2.5) measurements support this interpretation because the short-lived PM2.5 has similar pattern on timescales shorter than a few days, but it has much less monthly and longer timescale variations compared to $CO_2$.

Why is the COVID signal relatively clear in Chengdu, but not in Beijing? The answer lies in the differences in weather. Chengdu, situated within the Sichuan Basin in southwestern China, is surrounded by great mountain ranges including the Tibetan Plateau and has generally very calm weather with a famously known atmospheric inversion layer that is rarely broken[17], whereas Beijing, being at the edge of the North China Plain is subject to large weather fluctuations, frontal passages and seasonal shifts. Thus, while weather tends to dominate in Beijing, its variation is sufficiently small in Chengdu to allow COVID signal to be revealed.

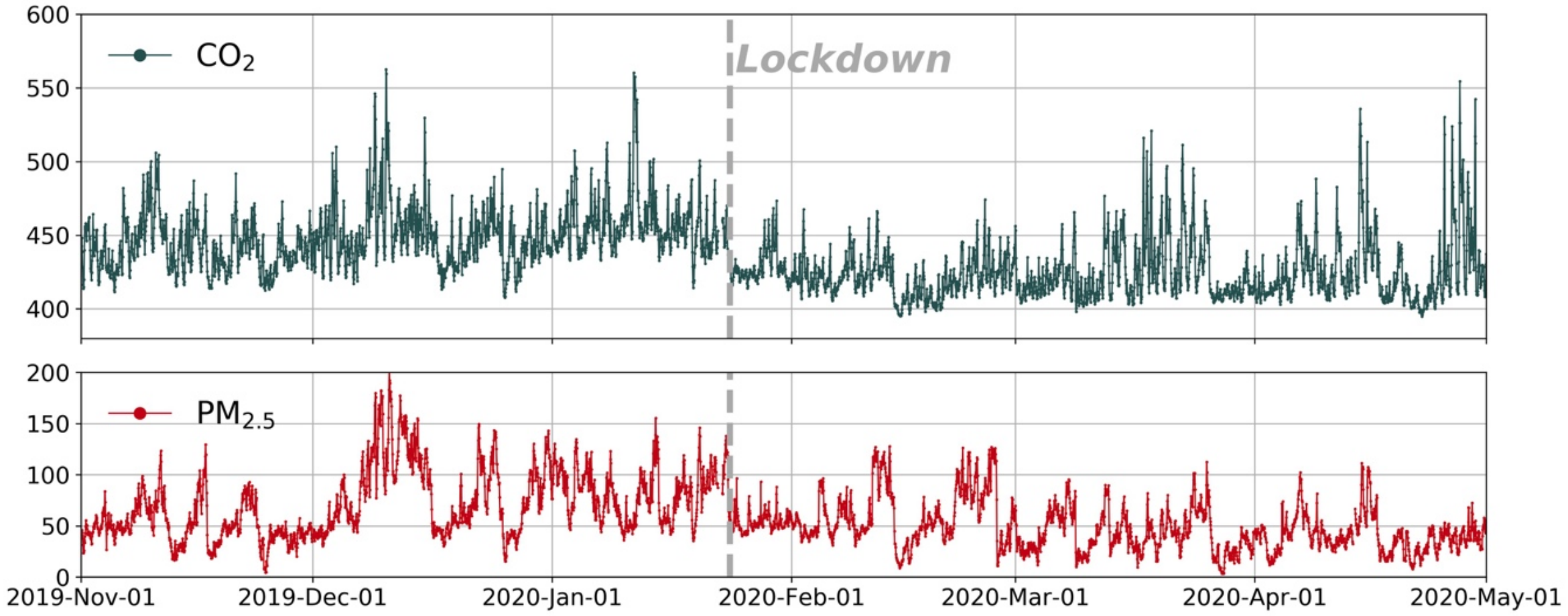

*Figure 7. Hourly $CO_2$ and PM2.5 measured in Chengdu, a city in the Sichuan Basin to the east of Tibet plateau. An abrupt drop on Jan 24, 2020, following the Lunar New Year and city-wide lockdown is clearly visible.*

**Direct observation of on-road $CO_2$ concentration** has been conducted in Beijing periodically since 2017 using mobile platforms[18,19] (See Method). Some of these '$CO_2$-tracking' trips took advantage of light-weight low-cost $CO_2$ sensors[20]. Three trips were selected from before, during, and after the COVID-19 lockdown while minimizing differences in other factors such as weather condition, rush hour, and weekend effects. The average $CO_2$ is 513 ppm before, 455 ppm during, and 501 after the height of lockdown (Fig. 8). To further remove the compounding effect of still somewhat differing weather conditions, we subtracted $CO_2$ concentration measured at the IAP tower station. This difference can be thought as a traffic-induced on-road '$CO_2$ enhancement' relative to a 'city background'. This 'traffic enhancement' is 65, 30 and 50 ppm respectively for the three periods. The more than 30 ppm less traffic $CO_2$ during COVID, and still somewhat depressed value during recovery is consistent with direct traffic data, not surprisingly, because the reduced transportation is the largest source of $CO_2$ reduction during lockdown in cities.

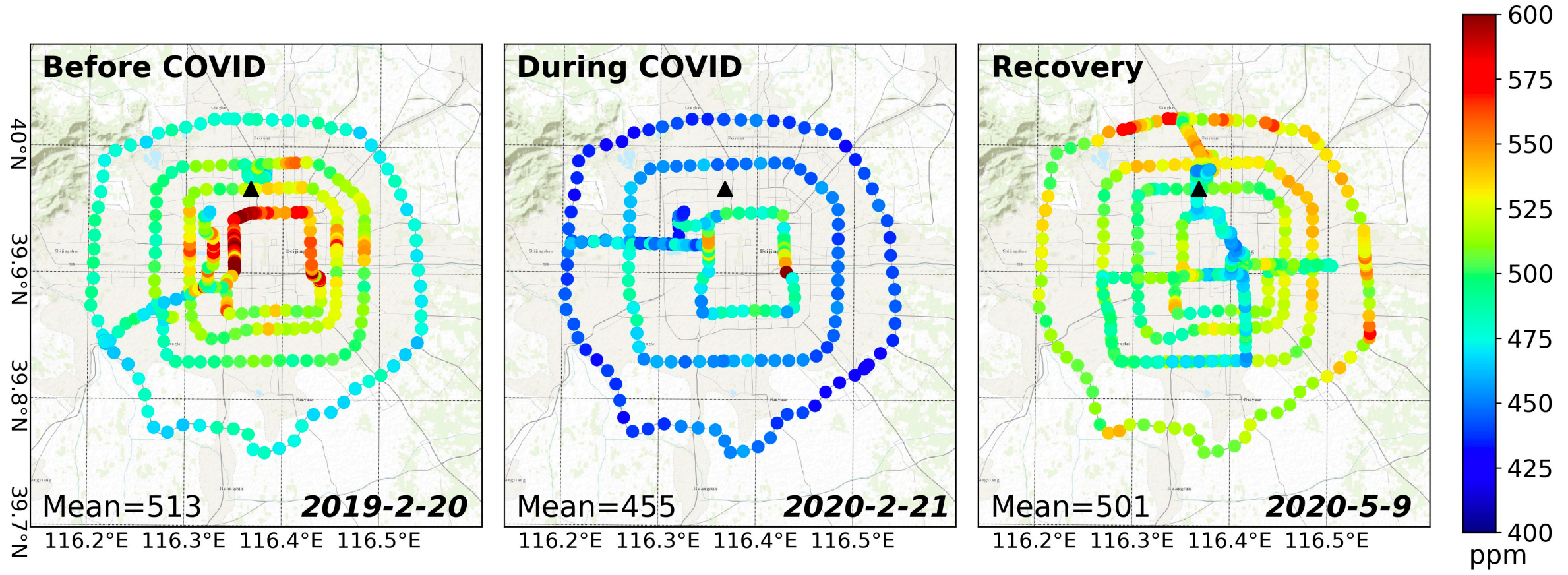

*Figure 8. On-road $CO_2$ concentration measured before, during and after COVID-19 lockdown in Beijing. These '$CO_2$ tracking' car trips covered the city's 4 major transportation arteries, from the Second (innermost) to the Fifth Ring Road. The west-east distance of the Fifth Ring Road is about 30 km. Each dot is 1-minute average of the original 1s or 2s data. Station data from the IAP tower, marked by black triangle, is used as 'city background' to compute on-road enhancement in the text.*

## Discussion

The detectability of COVID-19 $CO_2$ signal depends strongly on spatial and temporal scales. Our results show that the carbon satellite GOSAT is able to detect a short-term global mean $CO_2$ anomaly decrease of 0.2-0.3 ppm, a number below the satellite instrument's targeted accuracy of individual measurements. This somewhat surprising result is made possible by spatial and temporal averaging to compensate for the sampling sparsity, and importantly also by the cancellation of biosphere and weather variability at large-scale. As a result, the drop in global-scale $CO_2$ anomalies was dominated by the spatially coherent COVID emissions reduction, followed by a biospheric recovery from previous year's relative carbon loss, which became more important in the spring season, while weather impact fluctuated. One implication for detection is the need for meticulous approaches in enhancing signal-to-noise ratio and maximizing spatial-temporal data coverage[21,22]. A critical perspective here is the focus on sub-annual time scale which has received little attention in the past compared to the much larger $CO_2$ seasonal cycle and interannual variability. Our results suggest that current observation and modeling capabilities can depict sub-annual variations with some consistency, and the COVID period has the largest sub-annual $CO_2$ anomaly in the last 10 years.

Surface observations of the atmospheric background $CO_2$ concentration have variabilities up to 1-2 ppm, substantially larger than the ~0.2 ppm COVID fossil signal. However, the modeling results are sufficiently realistic in capturing sub-annual variabilities consistent with station observations, lending support in using model experiments to separate the COVID effect from the larger weather and biospheric variability.

At urban scale in many cities such as Beijing, New York City and Paris, atmospheric variability dominates the signal. Seasonal variations in weather patterns prevented us from discerning the COVID signal with confidence despite of the large fossil fuel emissions changes expected there. A major caveat is that the model resolution is too coarse to resolve the cities properly and the real signal is likely stronger than seen here and it would be better simulated by meso-scale models. Moreover, where weather variability is modest such as in Chengdu, the lockdown caused a clear $CO_2$ reduction. Additionally, decrease in on-road $CO_2$ enhancement larger than 20 ppm in Beijing was observed which is perhaps the most direct observation of localized emissions reduction.

Despite of the dramatic reduction in economic activities during 2020 COVID-19 worldwide lockdown, the short-duration of the event has left only a small signature in the atmospheric $CO_2$ which results from fossil fuel emissions accumulated over two hundred years due to $CO_2$'s long life-time in the atmosphere. Nevertheless, our analysis demonstrates that its global-to-local impacts are already detectable, albeit still imprecisely, by current carbon monitoring systems using a variety of approaches, and that meaningful causality attribution to fossil fuel, biosphere and weather can be made by combining model and observations. Continued improvement and expansion of such capabilities can play a critical role in monitoring and verification of fossil fuel emissions reduction target at local, country, and global scales. They can also facilitate climate mitigation efforts from governments, cities, institutions and citizens.

**Acknowledgements**
We are grateful to the providers of the following datasets, made available in a timely fashion: the GOSAT satellite data by JAXA and NIES, the NOAA GLOBALVIEW-CO2 ObsPack data via Andy Jacobson, Kirk Thoning and Pieter Tans, the NASA GISS temperature data and NOAA CPC PREL/OPI data. We thank the students and volunteers involved in the Beijing carbon monitoring project, in particular the $CO_2$-tracking trips. We thank discussion with Ruqi Yang, Wenhan Tang, and David Crisp on detecting COVID signal, and Anna Karion, James Whetstone, Israel Lopez-Coto on sensor development and regional $CO_2$ modeling. We acknowledge support from MOST (2017YFB0504000), NOAA (NA18OAR4310266), and NIST (70NANB14H333).

**Author contribution**
NZ, PH, DL and RD designed the study. NZ and ZQL designed and ZQL conducted the model simulations. CM, DL, RD, and ZQL did the model development. QC conducted the biosphere model simulations. ZL, TO, ZQL, NZ, PH produced the fossil fuel emissions data. TO and SM provided GOSAT interpretation. PH, BY, PW, WS, NZ, and DL organized the observations and CO2-Tracking trips in Beijing. NZ, ZQL and DL led the analysis and wrote the paper, and all participated in synthesis and helped writing.

## Competing interests
The authors declare no competing interests.

# Method

## Atmospheric transport model simulation

To simulate the atmospheric $CO_2$, the model solves the carbon mass balance equation:

$$\frac{dCO_2}{dt} = F_{net} \equiv F_{FE} + F_{TA} + F_{OA} \qquad (1)$$

where $\frac{dCO_2}{dt}$ is the atmospheric $CO_2$ growth rate, $F_{FE}$ is fossil fuel emissions, $F_{TA}$ is terrestrial to atmosphere flux, $F_{OA}$ is ocean to atmosphere flux, and $F_{net}$ is net surface to atmosphere flux. The model is run in a 'forward' fashion for each 3-dimensional model grid point (location), forced by the three fluxes, as well as meteorological variables for atmospheric transport and mixing. Here we use GEOS-Chem atmosphere transport model v12.5.0 (http://acmg.seas.harvard.edu/geos/) at a horizontal resolution of $4^o \times 5^o$ with 47 vertical levels. The fluxes (below) at different resolutions are re-gridded to $4^o \times 5^o$. The model is driven by the meteorology field MERRA2 from the NASA Global Modeling and Assimilation Office. Details of setup and evaluation were described earlier[23,24]. The simulation period was from 1 January 2008 to 31 May 2020, and data after January 2010 were used for analysis.

## Fossil fuel $CO_2$ emissions

$F_{FE}$ combines a number of sources, including the Global Carbon Project (GCP) annual country-level carbon budget for 1959-2018 [25] with update for 2019 by Le Quere et al. (2020) [1], daily data for 2019-2020 from Liu et al. (2020)[2], the TIMES hourly scaling factor of Nassar et al. (2013) [26], and the spatial information of the ODIAC database of Oda et al. (2018) [14].

Recently, a daily resolution, country-level data became available [2]. This novel dataset achieved daily resolution by taking advantage of a variety of sector-based energy and economic activity statistics, including real-time traffic data and daily electricity generation data of major power suppliers. However, the Liu et al. (2020) data was available only for early 2019 and 2020, and for this work, it was updated to cover the period of Jan 2019 through May 2020. For the years 2008-2018 when we do not have daily resolution data, we use the 2019's daily variation as a surrogate but retain their annual total. Because the emissions of Liu et al. (2020) for 2019-2020 are slightly different from GCP, in order to maintain the consistency of $F_{FE}$ from 2018 to 2019-2020, we use the country-level GCP fossil fuel values as a constrain to rescale the yearly total $F_{FE}$ from 2008 to 2020, and the same scaling factor for 2019 is used for 2020 to obtain a harmonized time series. Further, we combine the diurnal scaling factor from the TIMES method of Nassar et al.[26] and the daily national $CO_2$ emissions of 2019 of Liu et al. [2] to obtain an hourly country-level $CO_2$ emission in 2019 and 2020.

The gridded spatial information comes from the Open source Data Inventory of Anthropogenic CO2 emission (ODIAC) [14]. ODIAC uses the annual country-level fuel consumption bases $CO_2$ emissions estimates [27] and disaggregate to 1 km or 1 degree resolution using satellite night light observations and point source data. The annual data was disaggregated to monthly based on a climatological seasonal cycle. Here we simply disaggregated the country-level hourly data above to 1°×1° with the spatial information of ODIAC. Since COVID-19 reduction is more concentrated in cities with major reduction in transportation[2,28], this disaggregation in proportion to ODIAC's spatial pattern may underestimate the reduction in metropolitan regions.

Altogether, the method can be summarized in the following equation for 2008~2018:

$$F_{FE}{}^{c,i,j,y,t_h} = ODIAC^{c,i,j,y} \times \frac{GCP_{tot}{}^{c,y}}{ODIAC_{tot}{}^{c,y}} \times \frac{LZ^{c,2019,t_d}}{LZ_{tot}{}^{c,2019}} \times TIMES^{i,j,t_{diurnal}}$$

and for 2019~2020:

$$F_{FE}{}^{c,i,j,y,t_h} = ODIAC^{c,i,j,2018} \times \frac{GCP_{tot}{}^{c,2019}}{ODIAC_{tot}{}^{c,2018}} \times \frac{LZ^{c,y,t_d}}{LZ_{tot}{}^{c,2019}} \times TIMES^{i,j,t_{diurnal}}$$

where y is year, $t_h$ is hour, $t_d$ is day, $t_{diurnal}$ is the diurnal cycle. c is country, i is longitude, j is latitude, tot is the yearly total value. $F_{FE}^{c,i,j,y,t_h}$ is the emission of country c at location i, j at time $t_h$ of the year i. The four datasets used to obtain this harmonized labeled ODIAC, GCP, TIMES and LZ, with LZ for our updated Liu et al. (2020) dataset.

**The terrestrial biospheric flux**

$F_{TA}$ is simulated by a dynamic vegetation and terrestrial carbon cycle model VEGAS [10,15]. The model is forced by observed climate variables including monthly precipitation, hourly temperature and radiation, and historical land use pattern as well as atmospheric $CO_2$. The model was run at hourly time step and 0.5°×0.5° resolution from 1901 to Apr 2020. This version 2.6 of VEGAS follows largely the simulation protocol of the TRENDY [29] and the MsTMIP [30] terrestrial model intercomparison projects with some model and near real time (NRT) forcing data updates. Carbon cycle models have been applied to long-term, interannual and seasonal variations extensively, but rarely in sub-annual changes of interest here. VEGAS has been shown to be among the models better at simulating such short-term changes [31,32].

**The ocean-atmosphere carbon flux**

$F_{OA}$ uses the spatial pattern of $pCO_2$ observation derived fluxes of Takahashi et al. (2009)[33]. To obtain the temporal variation, we rescaled the Takahashi spatial pattern for the year 2013 with the temporal evolution of $F_{OA}$ from the GCP annual carbon budget analysis which is based on estimates from multiple ocean carbon cycle models [25]. The carbon budget was up to only 2018. For the year of 2019 and 2020, the GCP ocean values are linearly extrapolated using the values from the previous 10 years. The annual carbon budget thus does not contain possible sub-annual contribution from ocean, which is generally believed to be small compared to land and fossil flux anomalies.

**Model sensitivity experiments**

To delineate contribution to $CO_2$ changes from fossil fuel emissions $F_{FE}$, biospheric flux $F_{TA}$, and weather, we designed three sets of experiments:

1. BWC (Biosphere+Weather+COVID): the full experiment described above with realistically varying biospheric fluxes $F_{TA}$, weather, and $F_{FE}$ including COVID-induced emissions reduction.
2. BW (Biosphere+Weather, no COVID): as in BWC, but replace $F_{FE}$ of 2020 with that of 2019.
3. B (Biosphere only): as in BW, but replace all years' meteorology (wind, etc.) with that of 2019.

Thus, compared to BWC, Experiment BW removes the effect of COVID emissions, while Experiment B further removes weather effect. The differences among these experiments show the effect on $CO_2$ of each individual factor. In Fig. 4, anomaly for experiment B is calculated as detrended anomaly, while those of BW and BWC represent the differences between 2020 and 2019, following the experiment design.

**$CO_2$ observations: satellite column $CO_2$ from GOSAT**

The satellite column $CO_2$ data is Level-3 product from the National Institute for Environmental Studies (NIES) [22] (http://www.gosat.nies.go.jp/en/about_5_products.html). The L3 data derived from the L2 data with spatial interpolation using Kriging technique. The data is at 2.5 °×2.5 ° resolution, available from August 2009 to April 2020. There are 3 months of missing data (December 2014, January 2015 and December 2018), which were gap-filled with spline method. There are known biases in oceanic glint data, so we only used land data in our analysis. For this reason, regional average analyses in such as Fig. 2b are over land only for both model and GOSAT to facilitate comparison. Additional uncertainty comes from missing GOSAT data in regions such as the core of the Amazon due to persistent cloud cover and the northern boundaries where large solar zenith angle may lead to larger uncertainty (e.g., Fig. 15 of reference [22]).

**Global network of surface $CO_2$ observations**

The surface station data are from the ObsPack framework [16] that collects a great variety and numbers of in-situ, flask sampling, aircraft and other $CO_2$ measurements. The five stations data used in our analysis are all flask sampling data. These are baseline stations managed by NOAA that have been in operation for several decades. Great care is taken to sample air representative of large-scale atmospheric background condition. Data product used here is GLOBALVIEW+5.0, with most recent update ObsPack NRT V5.0 provided by NOAA's CarbonTracker team[34,35]. The most recent year's data have been quality-controlled by an automated procedure, and they may still be subject to modifications from further manual quality control.

**City $CO_2$ station observations**

A network of 6 tower stations using high accuracy Picarro $CO_2$ analyzers has been running since 2018 as part of the Beijing-Tianjin-Hebei (JJJ) carbon monitoring project [19], run by the Chinese Meteorological Administration (CMA) and the Institute of Atmospheric Physics (IAP) of the Chinese Academy of Sciences. The $CO_2$ analyzers are calibrated 4 times a day, with calibration gas tracing to World Meteorological Organization (WMO) standard. The data have a nominal accuracy of 0.1 ppm. A network of low-cost $CO_2$ sensors has been running in various stages of development since 2016, as a collaborative effort between IAP, the University of Maryland, and the US National Institute for Standards and Technology (NIST). These sensors were found to be able to achieve an accuracy of ~5 ppm after calibration and environmental correction[20]. The data used in this paper for Beijing stations are measured with Picarros while the data in Chengdu was from a low-cost sensor node with 3 individual $CO_2$ sensors.

**On-road $CO_2$ observation in Beijing before, during and after COVID-19 lockdown**

We conducted several on-road $CO_2$ measurements in Beijing and surrounding area using mobile platforms before, during and after COVID-19 lockdown. Because urban $CO_2$ concentration is strongly influenced by weather, we selected three trips with closest weather as possible for the days of February 20 of 2019, February 21 of 2020, May 9 of 2020. Additionally, we calculate the on-road $CO_2$ enhancement relative to a city 'background' measured at the IAP tower station. We used $CO_2$ sensors of different accuracy including Picarro and LI-COR LI-7810, both mounted inside car with air inlet from above roof [18]. We also used low-cost sensors mounted on windshield, calibrated before and after each trip[19]. Some of the sensors were in the same car with Picarro and their agreement was found to be within 5 ppm.

**Analysis of $CO_2$: separate sub-annual anomalies from trend and seasonal cycle**

Atmospheric $CO_2$ data contains variability on a variety of time scales, from long-term increasing trend driven by fossil fuel emissions and carbon sinks [36], Decadal variations [37], interannual variability dominated by ENSO [10], to a prominent seasonal cycle [15,38] in response to the annual growth and decay of the biosphere. The possible COVID-19 signal of interest here lasts for few months on sub-annual (month to intra-seasonal) timescales. Monthly-scale high-frequency variabilities are generally less well studied and are often filtered out so as to focus on seasonal and longer-term changes [39].

Here we calculate sub-annual anomalies using a 4-step 'detrended anomaly' approach termed DCA (Detrending, Climatology, Anomaly):

1) A 12-month running mean is applied to the original $CO_2$ data. The running mean contains mostly signals longer than a year, including long-term trend, interannual to decadal variations (Fig. 9a)
2) This running mean is then subtracted from the original $CO_2$ data. The result is considered 'detrended' and is dominated by seasonal cycle (Fig. 9b black line)
3) A climatology is then calculated as the mean seasonal cycle (Fig. 9b green line).
4) The sub-annual anomalies (detrended anomalies) is the difference between the detrended $CO_2$ and its climatology

The approach using running mean to remove low-frequency signals has been used by various authors, e.g., to study the $CO_2$ seasonal amplitude change [15]. The last two steps consist of a standard definition of climatology and anomaly. In comparison to the DCA method, the standard climatology/anomaly method (simply called CA method here) does not involve detrending, and it retains low-frequency signal. Therefore, the DCA method is suitable for finding $CO_2$ sub-annual anomalies, while the CA method is used for flux analysis such as in Fig. 2a which contains both interannual and sub-seasonal information.

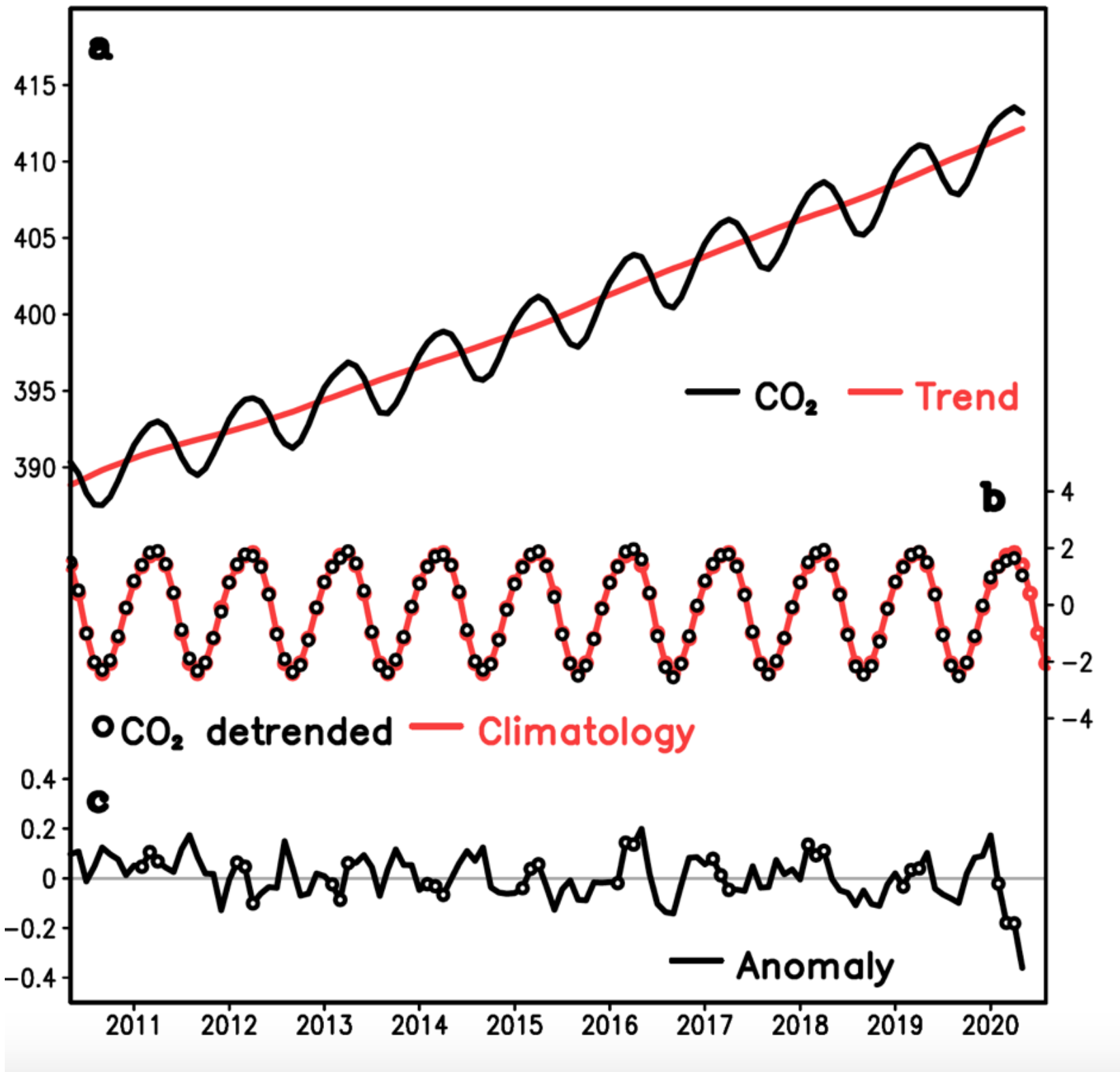

*Figure 9. The DCA (Detrending, Climatology, and Anomaly) method for finding sub-annual anomalies in a typical $CO_2$ time series (shown are model simulated global mean $CO_2$). (a) Original $CO_2$ data (black) and 12-month running mean (red); (b) $CO_2$ detrended (black) and its climatology (red); (c) Detrended anomalies ($CO_2$ detrended minus its climatology), with open circles marking Feb-April of each year.*

## References for Method